\documentclass [prl, superscriptaddress, showpacs, twocolumn]{revtex4}

\usepackage{graphicx}
\usepackage{dcolumn}

\newcommand{\grad}{^\circ}
\newcommand{\prca}{Pr$_{1-x}$Ca$_x$MnO$_3$}
\newcommand{\rdmno}{R$_{1-x}$D$_x$MnO$_3$}
\newcommand{\prcaq}{Pr$_{0.60}$Ca$_{0.40}$MnO$_3$}

\newcommand{\lacah}{La$_{1/2}$Ca$_{1/2}$MnO$_3$}
\newcommand{\bisrh}{Bi$_{1/2}$Sr$_{1/2}$MnO$_3$}
\newcommand{\rcamno}{R$_{1/2}$Ca$_{1/2}$MnO$_3$}
\newcommand{\rdmnoh}{R$_{1/2}$D$_{1/2}$MnO$_3$}

\newcommand{\mnt}{Mn$^{3+}$}
\newcommand{\mnq}{Mn$^{4+}$}
\newcommand{\tco}{$T_{CO}$}
\newcommand{\tn}{$T_N$}

\newcommand{\dmno}{$d_{MnO}$}
\newcommand{\mnomn}{Mn$_1$-O$_3$-Mn$_2$}
\newcommand{\distor}{$\Delta(\times10^{-4}$)}

\begin{document}

\title{Single crystal structural study of the charge ordered phase of Pr0.6Ca0.4MnO3:
electronic localization beyond the atomic level.}

\author{ A. Daoud-Aladine}
\affiliation{Laboratoire L\'{e}on Brillouin, CEA-CNRS Saclay,91191 Gif sur Yvette}
\author{J. Rodr{\'{\i}}guez-Carvajal}
\affiliation{Laboratoire L\'{e}on Brillouin, CEA-CNRS Saclay,91191 Gif sur Yvette}
\author{L. Pinsard-Gaudart}
\affiliation{Laboratoire L\'{e}on Brillouin, CEA-CNRS Saclay,91191 Gif sur Yvette}
\affiliation{Laboratoire de Physico-Chimie de l'Etat Solide, Universit\'{e} Paris Sud, B\^{a}t. 414, 91405 Orsay}
\author{M.T. Fern{\'a}ndez-D{\'{\i}}az}
\affiliation{Institut Laue Langevin, 38042 Grenoble}
\author{A. Revcolevschi}
\affiliation{Laboratoire de Physico-Chimie de l'Etat Solide, Universit\'{e} Paris Sud, B\^{a}t. 414, 91405 Orsay}

\date{\today}

\begin{abstract}
In the \rdmno \ manganites (R: rare earth, D: divalent cation), the structural phase transition at
\tco \ is commonly interpreted as a concomitant charge and orbital ordering (CO/OO) process driven by
a cooperative Jahn-Teller effect and Coulomb repulsion forces. Full crystal structure refinement, from
a neutron diffraction experiment below \tco \ on a \prcaq \ single crystal, gives us a model for the
displacement of atoms with respect to the high temperature (HT) phase that invalidates the standard
model based on the CO/OO picture. An alternative picture is that the phase transition and charge
localization at \tco \ arise from a spatially ordered condensation of ferromagnetic Mn-Mn pairs, in
which Mn atoms keeps an intermediate valence state.
\end{abstract}


\maketitle

\begin{table}[b]
\caption{\small Octahedral distortions in \prcaq.
$\Delta=\frac{1}{6}\sum_{i=1}^{6}{|\frac{d_{MnO_{i}}-<d_{MnO}>}{<d_{MnO}>}|^{2}}$ measure the
octahedron regularity. The Bond Valence Sums (BVS), uses the \mnt d$_0$-parameter for all sites (see
ref\cite{Rodriguez98a} for definitions).} \label{mnodis}
\begin{ruledtabular}
\begin{tabular}{ccc}
$T=280K$   &\multicolumn{2}{c}{$T=195K$}\\ \cline{1-1} \cline{2-3} Mn site &Mn$_1$ site &Mn$_2$ site\\
Mn-O$_{ap}$:1.953(1)$(\times2)$ &Mn$_1$-O$_{1}^{'}$:1.951(3) &Mn$_2$-O$_{3}^{'}$:1.957(3)\\
Mn-O$_{eq}$:1.967(2)$(\times2)$ &Mn$_1$-O$_{2}^{'}$:1.942(3) &Mn$_2$-O$_{4}^{'}$:1.940(3)\\
Mn-O$_{eq}$:1.953(1)$(\times2)$ &Mn$_1$-O$_{1}$:1.879(2) &Mn$_2$-O$_{4}$:1.899(2) \\
                                &Mn$_1$-O$_{1}$:2.053(2)        &Mn$_2$-O$_{4}$:2.028(2) \\
                                &Mn$_1$-O$_{3}$:1.980(2)        &Mn$_2$-O$_{3}$:2.011(2) \\
                                &Mn$_1$-O$_{2}$:1.955(2)        &Mn$_2$-O$_{2}$:1.899(2) \\
\dmno=1.957(1) \AA              &\dmno=1.960(1) \AA             &\dmno=1.955(1) \AA\\ BVS=3.52
&BVS=3.53(1)                    &BVS=3.50(1) \\ \distor=0.12                    &\distor=6.99
&\distor=6.506\\
\end{tabular}
\end{ruledtabular}
\end{table}

\begin{figure*}[btp]
\begin{center}
\includegraphics* [width=220pt] {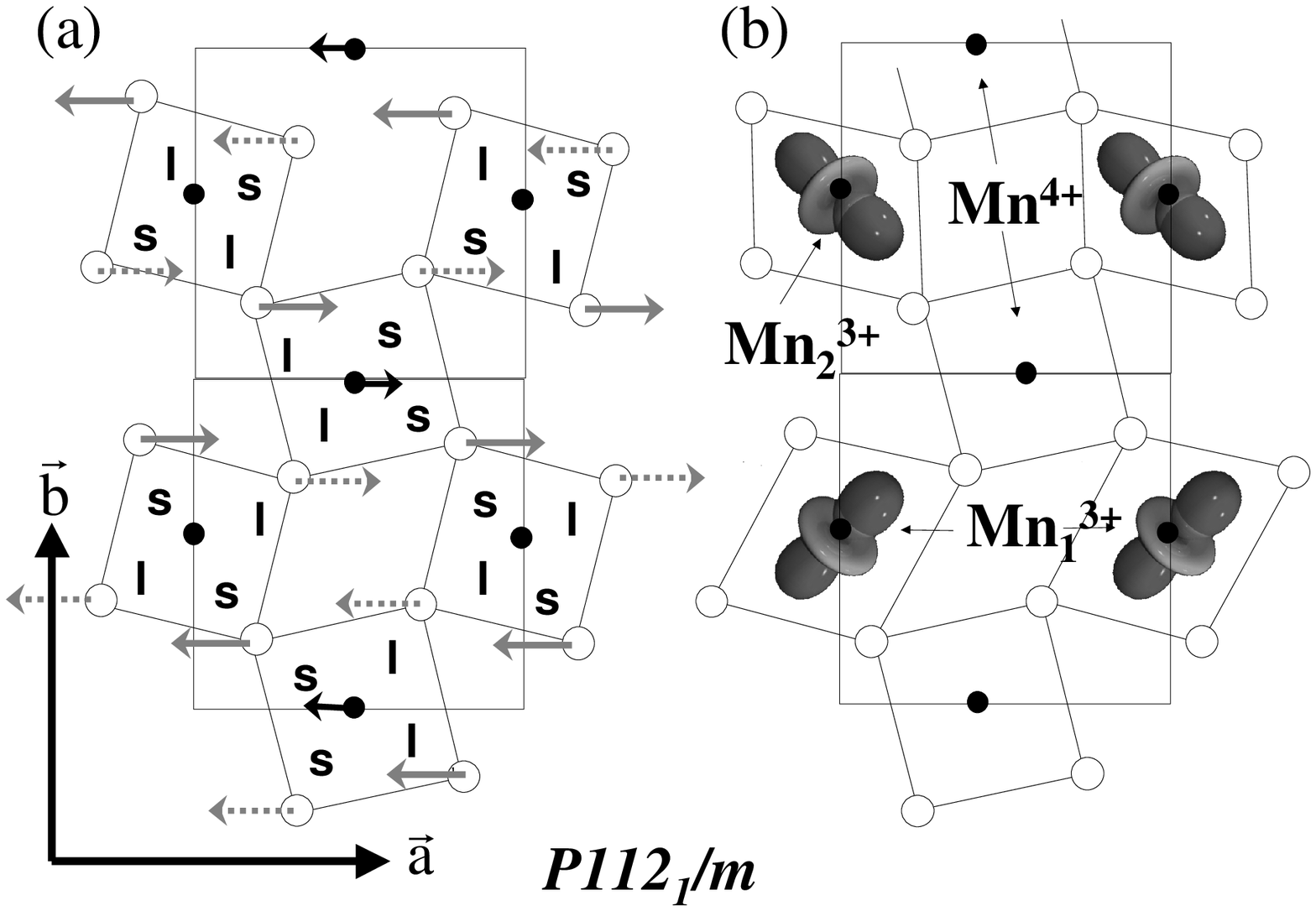} \ 
\includegraphics* [height=160pt] {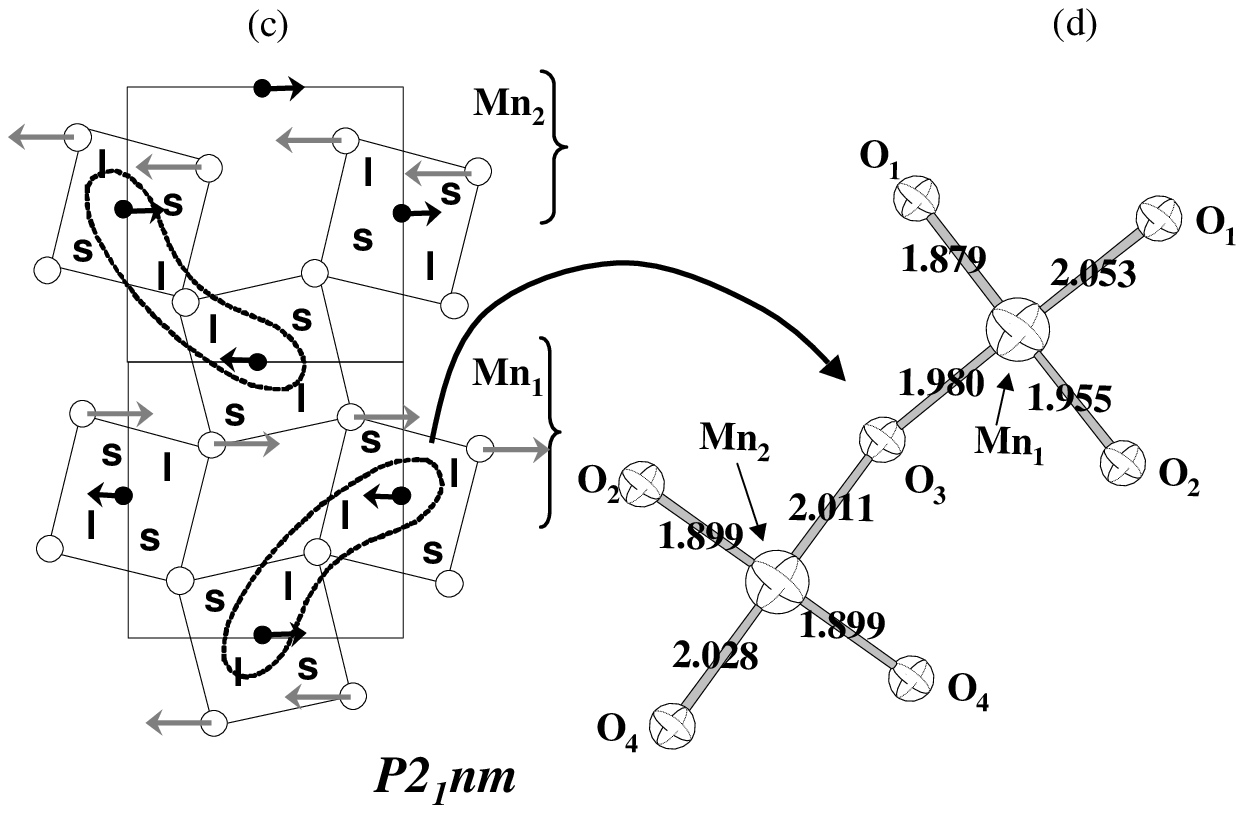}
\caption{\small MnO$_2$ (a,b) planes of the structure: schematic representation of the displacements
of the atoms with respect to their average $Pbnm$ positions, that are exaggerated for clarity, in the
$P112_1/m$ symmetry (a) and  with $P2_1nm$ symmetry (c) respectively. Labels "l"  and "s" are for long
($\approx$2 \AA) and short ($\approx$1.9 \AA) Mn-O distances. The refinement of from our single
crystal data with the  $P112_1/m$ symmetry the gives: for the 827 measured fundamental lines,
R$^f_{F^2}$ factor : 4.21\% and for 722 superstructure lines, R$^{ss}_{F^2}$ factor : 39.3\%. With the
$P2_1nm$ symmetry refinements gives R$^f_{F^{2}}$ factor : 2.62\% and R$^{ss}_{F^{2}}$ factor : 10.8
\% (see Fig.\ref{refineODC}) \label{p21surm}. (b) is the standard CO/OO model deduced from (a). (d)
show the Mn-O distances around one Mn$_1$-O$_3$-Mn$_2$ unit, represented by the lobes on the figure
(c).} \label{models}
\end{center}
\end{figure*}

The calcium-doped manganites \rcamno ($x\approx0.5$) (R: rare-earth) are known to display a structural
phase transition at \tco \ attributed to charge and orbital orderings, because it is associated with a
jump in the resistivity and to the onset of superstructure reflections in the low temperature (LT)
phase with $\mathbf{q}=(0,1/2,0)$ in electron, X-ray and neutron diffraction patterns \cite{Mori98,
Jirak85}. Moreover, they display another unique signature consisting of a complex AF CE-type spin
ordering at a temperature \tn$<$\tco. This can be comprehensively understood using the semi-empirical
Goodenough-Kanamori-Anderson (GKA) rules for the superexchange interactions, applied to a model of
orbital ordering of the \mnt \ $d_{z^2}$ orbitals in the (a,b) plane, proposed in the late fifties by
Goodenough \cite{Goodenough55, Wollan55}. The fact that Goodenough's model (GM) suggests a charge
ordered pattern of \mnt \ and \mnq \ ions in a NaCl-type lattice, has raised the idea that the
magneto-transport properties of these compounds are strongly influenced by dynamic CO/OO driven by the
combined effect of Coulomb repulsion and Jahn-Teller distortion around the \mnt \ sites \cite{Chen99a,
Radaelli97c}. The superstructure is the result of modulated atomic displacements accompanying the
electronic localization, but the very nature of the transition should be deduced from the analysis of
a detailed structure determination not needing \emph{a priori} constraints. It is expected that CO
will result in the setting up of MnO$_6$ octahedra of different average Mn-O distances (\dmno), as it
is empirically well established by the Bond Valence Model (BVM) \cite{Brown92}. Note that in LaMnO$_3$
and CaMnO$_3$, the \dmno \ are respectively 2.02 {\AA} (\mnt) and 1.90 {\AA} (\mnq). A NaCl-like CO without OO
is also compatible with a transition characterized by $\mathbf{q}_{CO}=(0,0,0)$ in the $Pbnm$ setting,
so that several subgroups of $Pbnm$ are possible. This is the case of the charge disproportionation
observed in the metal-insulator transition of RNiO$_3$ (R=Ho, Y, Er, Tm, Yb, and Lu) \cite{Alonso00}.
A Jahn-Teller driven orbital ordering should also differentiate the octahedra by their distortion. The
experimental determination of these displacements has been attempted on several manganites, following
the work of Radaelli et al. \cite{Radaelli97c} on \lacah. The small number of weak superstructure
reflections observable by X-ray or neutron powder diffraction makes the published results dubious,
since structural models have to be constrained in order to obtain stable refinements
\cite{Radaelli97c, Blasco97, Woodward99, Jirak00, Richard99}. The constraints that have been used are
those one would expect from the \emph{a priori} assumption that the GM is the \emph{correct} one. The
GM of CO/OO orbital ordering for \rdmnoh \ perovskites have only been justified, from powder
diffraction experiments, by introducing constraints between the free parameters of the space group
$P112_1/m$ that imposes a pattern of atom displacements
as depicted in Fig.\ref{models}a. 

By contrast, in LiMn$_2$O$_4$ \cite{Rodriguez98a}, a Mn oxide with a cubic spinel structure, charge
ordering, together with a cooperative Jahn-Teller effect, is clearly established below the structural
transition around RT. The CO/OO picture is so widely accepted to describe the properties of half doped
manganites that the use of the associated jargon hides the today's lack of a non-ambiguous atom
displacements pattern properly explaining the intensity of a high number of superstructure
reflections. The literature in the field is plenty of assumptions that are not justified from our
point of view. The appearance of reflections of type $(0,k,0)$ with $k$ odd, in $Pbnm$ notation, is
automatically attributed to CO and the reflections indexed with the propagation vector
$\mathbf{q}=(0,1/2,0)$, $(h,k+\frac{1}{2},l)$, to OO. The appearance of these reflections are just due
to a structural phase transition, the origin of the phase transition may be other than CO/OO. The
literature concerning resonant x-ray scattering (RIXS) is particularly prone to these
misunderstandings \cite{von-Zimme99,Wakabayas00}, due to the fact that the interpretation of the
experimental outcomes is still controversial \cite{Joly00, Garcia01d}.

In this Letter, we report a complete structure refinement of a \prcaq \ single crystal. We selected
the Pr-Ca system for two reasons: (1) it is an archetype of the so called CO-manganites, in which the
CO-phase is known to be stable down to the $x=0.3$ doping level \cite{Jirak85,Jirak00,Tokura99}; (2)
the similar ionic radius of Pr$^{3+}$ and Ca$^{2+}$ makes the size mismatch effects negligible,
minimizing the differences between the average displacements, observed by analyzing the Bragg
reflections, and the local oxygen displacement induced by chemical disorder. The data collection was
performed using the 2D detector of the four circle diffractometer D9 at the Institute Laue-Langevin in
Grenoble. The study of the magnetic structure, at $T=15K$ (\tn$=170K$) gives a canted AF-CE structure
with an angle between the moment direction and the $\mathbf{c}$-axis equal to $\theta=28\grad$,
consistent with the dependence of $\theta$ as a function of the composition found in the literature.
The characterization of a crushed piece of the same crystal using synchrotron high resolution powder
diffraction (SHRPD), confirms the crystal is greatly homogeneous since it displays a single phase
pattern at $T=250K$ above \tco$=240K$. Its stoichiometry, refined by combined refinements of single
crystal neutron diffraction and synchrotron data is $x=0.40(1)$. Complementary XANES experiments have
also been carried out in other powder samples of similar compositions. A complete description of all
our synchrotron results, including the defect microstructure, as well as the technical details of the
neutron diffraction refinements is out of the scope of this Letter, and will be published elsewhere.

The problem of structure refinement of the neutron diffraction data in these orthorhombic perovskites
arises from the intrinsic twinning of the crystals. Here, it can be solved thanks to the small
distortion \prcaq \ present at all temperatures. A single integrated intensity has contributions from
different domains. This can be handled by the program FULLPROF \cite{Rodriguez93} that allows us to
refine a common structural model by adding the contribution of all domains. At $T=280K$, 416
reflections were used for the refinements, and the R$_{F^{2}}$ factor is 3.03\%. The structure is
orthorhombic with $Pbnm$ symmetry, very similar to the already published data \cite{Jirak85}. Its cell
parameters (SXHRDP data) are $a=5.4210(1)$\AA, $b=5.4460(1)$\AA \ and $c=7.6480(1)$\AA. It has a
unique Mn site and displays a pattern made of tilted nearly regular octahedra, whose \dmno \ is
characteristic of an intermediate valence site (see table \ref{mnodis}).

\begin{figure}[b]
\includegraphics* [height=110pt] {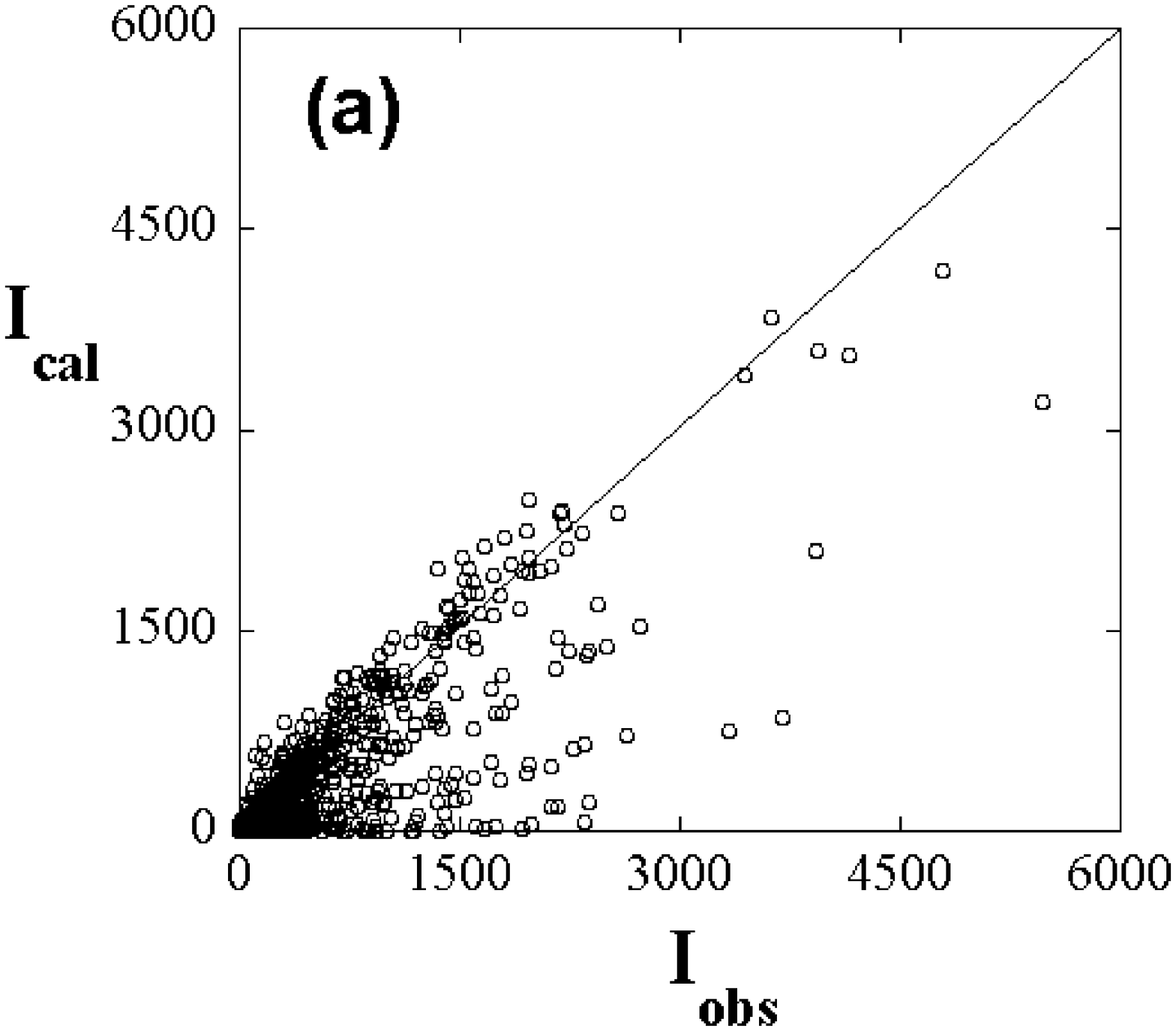}
\includegraphics* [height=110pt] {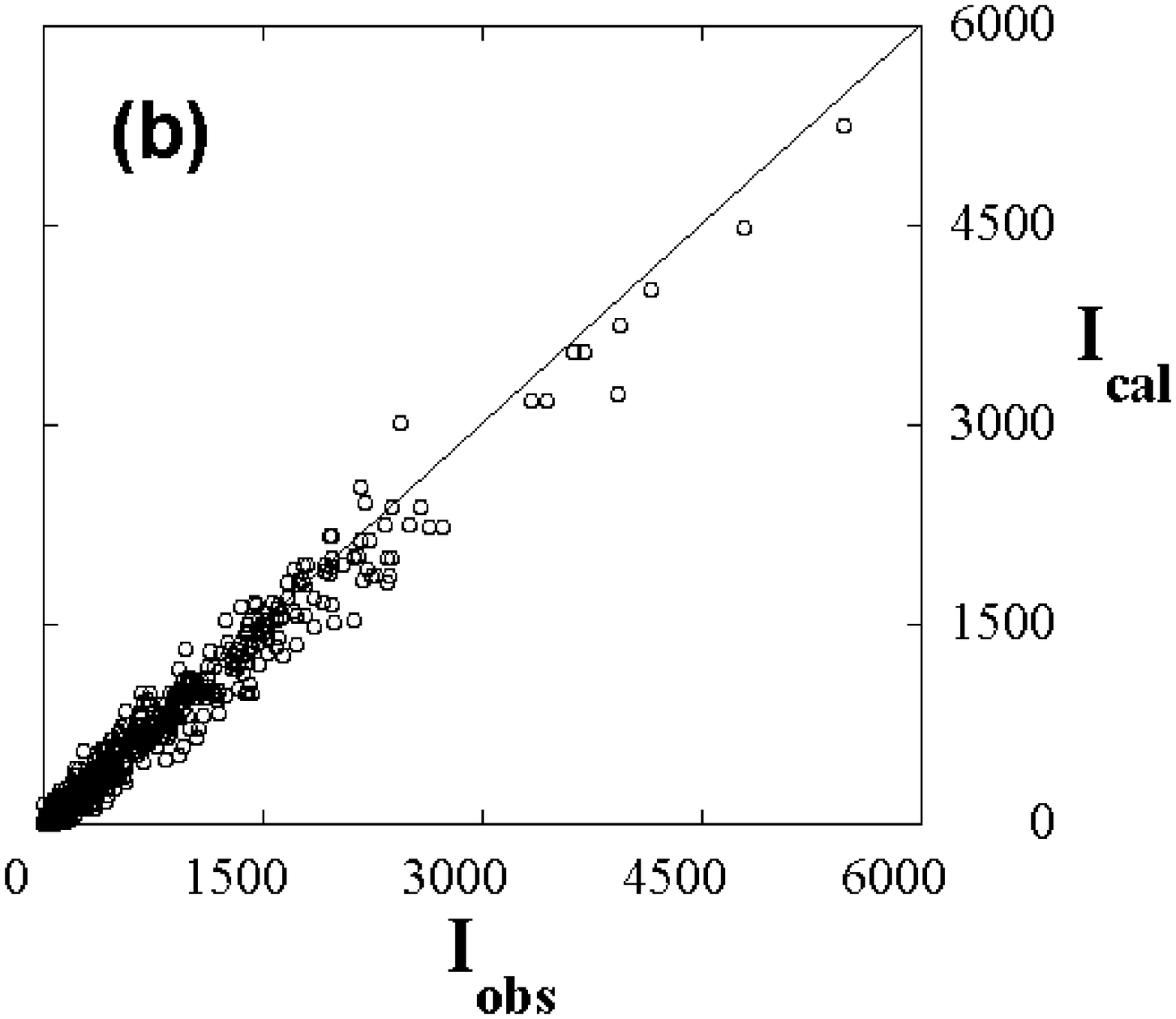}
\caption{\small Observed vs calculated intensity of superstructure lines indexed as $(h,k+1/2,l)$ in
the HT cell setting, in \prcaq \ with $P112_1/m$ (a) and  $P2_1nm$ symmetry (b) structural models.}
\label{refineODC}
\end{figure}

Below \tco \, at $T=195K$, the most intense superstructure peaks are found to have $2.7\%$ of the
intensity of the most intense fundamental peaks. Group theory gives six isotropy subgroups of $Pbnm$
having a doubled cell along $\mathbf{b}$, among which we have performed unconstrained refinements,
except for symmetry. Any one of the three isotropy subgroups keeping the $m$ mirror plane
perpendicular to the $\mathbf{c}$ direction of the HT phase, gives rather good results fits to the
fundamental reflections. But the intensity of superstructure lines are much better described by the
space groups $P2_1nm$ or $P11m$ than by the group $P112_1/m$ which, we recall, was the previously
reported candidate for the symmetry of the CO phase (see Fig. \ref{refineODC} and caption of
Fig.\ref{models} for the reliability factors). Note that $P11m$ being a subgroup of both $P112_1/m$
and $P2_1nm$, it could allow to describe displacements as expected from the GM. Our study demonstrates
that the experimentally refined displacement model with $P11m$ symmetry gives a strong $P2_1nm$
pseudo-symmetry. Indeed, the refined atomic positions give no significant differences using either
$P2_1nm$ or $P11m$ models, except that the number, and therefore the correlations, of the parameters
in $P11m$ are greater. Hence, the true space group is $P11m$, since we know from the SHRPD study, that
the metric is monoclinic, but the refined structural parameters can be described in the $P2_1nm$
group. They are given in Table \ref{str_par_195k} and a schematic pattern of the atomic displacements
is displayed in Fig.\ref{models}c. When we compare them to those of Fig.\ref{models}a, we see that in
both cases, the displacements are mainly along the $\mathbf{a}$-axis. The symmetry imposed inversion
center on the Mn sites identified to \mnt \ in the $P112_1/m$ description allow only symmetric oxygen
displacement around these Mn. This symmetry imposed constraint is lifted in the non-centrosymtric
group $P2_1nm$ (Fig. \ref{models}c). This allow all the Mn ions in the structure to move off center in
their slightly elongated MnO$_6$ octahedron which keep otherwise almost regular local orthogonal axes
(Fig. \ref{models}c). All these structural details are inconsistent with the ionic picture of
concomitant \mnt/\mnq \ CO and $d_{z^2}$ OO. In particular the persistence of an intermediate valence
state for the Mn atoms in the LT phase is confirmed by the very similar \dmno \ on the two sites (see
Table \ref{mnodis}).

\begin{table}[tbp]
\begin{center}
\caption{\small Refinement of structural parameters of Pr$_{0.60}$Ca$_{0.40}$MnO$_3$ at T=195K in the
orthorhombic $P2_1nm$ pseudo-symmetry, the real structure is monoclinic $P11m$ with cell parameters
$a=5.4315(1)$\AA, $b=10.8970(2)$\AA, $c=7.6370(1)$\AA, $\gamma=90.076(2)^o$. The doubled cell origin
is shifted by a (0 3/4 1/4) translation with respect to the high temperature orthorhombic $Pbnm$
phase. \label{str_par_195k} }
\begin{tabular}{lccccc}
\hline \hline
  Atom          & Wyck.& x       &y          &z          &B$_{iso}$\\
                & Pos. &         &           &           &(\AA$^{2}$)\\
\hline
 Pr/Ca$_1$      & 2a&  0.5121(4) & 0.8936(2) & 0         & 0.53(3)  \\
 Pr/Ca$_2$      & 2a&  0.4784(4) & 0.3614(2) & 1/2       & 0.77(3)  \\
 Pr/Ca$_3$      & 4b& -0.0023(4) & 0.1426(2) & 0         & 0.56(3)  \\
 Pr/Ca$_4$      & 4b&  0.9905(4) & 0.6088(2) & 1/2       & 0.25(2)  \\
 Mn$_1$         & 2a&  0         & 0.8756(2) & 0.2489(3) & 0.31(2)  \\
 Mn$_2$         & 2a&  0.9795(2) & 0.3746(1) & 0.7492(4) & 0.13(2)  \\
 O$_{1}$        & 4b&  0.3044(4) & 0.9845(1) & 0.2861(2) & 0.26(2)  \\
 O$_{2}$        & 4b&  0.7090(4) & 0.2676(1) & 0.7891(2) & 0.48(1)  \\
 O$_{3}$        & 4b&  0.2112(4) & 0.2328(1) & 0.7148(2) & 0.96(2)  \\
 O$_{4}$        & 4b&  0.7515(4) & 0.5191(1) & 0.2110(2) & 0.41(2)  \\
 O$_{1}^{'}$    & 2a&  0.4353(4) & 0.1125(2) & 0         & 0.41(2)  \\
 O$_{2}^{'}$    & 2a&  0.5743(4) & 0.1321(2) & 1/2       & 0.68(3)  \\
 O$_{3}^{'}$    & 2a&  0.0562(4) & 0.3758(1) & 0         & 0.33(2)  \\
 O$_{4}^{'}$    & 2a&  0.9104(4) & 0.3846(2) & 1/2       & 0.69(2)  \\
\hline \hline
\end{tabular}
\end{center}
\end{table}

The XANES study of the Mn valence state using the Mn K-edge we performed on a \prca \ ($x=1/2$) powder
sample lead to the same conclusion as the analysis recently published by Garcia et al.
\cite{Garcia01a}. The sample presents, at all temperatures, a spectra that have a sharp slope at the
edge. This contradicts the CO picture for the LT phase, where for a mixing of localized \mnt \ and
\mnq \ electronic states, the spectrum is expected to be linear combination of the end members
RMnO$_3$ and DMnO$_3$ XANES spectra, resulting in a broad shape and a pronounced structure. However,
the octahedral distortions below \tco \ manifest themselves as small changes that are located at
energies well above the edge. They reduce also the pre-peak intensity, which can be interpreted as
hole ordering on oxygen sites together with an increase in the covalency of Mn $3d$ and O $2p$ states
rather than \mnt/\mnq charge ordering \cite{Ignatov01}. The off-centering of Mn atom in our structural
model also make the dipole $1s-3d$ transitions allowed via $3d-4p$ mixing \emph{on} the absorbing
site, which had never been seriously considered since Mn sites have always been considered as
centrosymetric insofar \cite{Bridges01}. At last, this distortion leads also to an anisotropy of the
anomalous scattering factor, that may be sufficient to observe resonant "forbidden" reflections with
the RIXS technique \cite{Garcia01d}, which are not necessarily associated here, to different charge
densities around the Mn atoms. Therefore, this show that the apparent contradiction
\cite{Garcia01a,von-Zimme99} of the XANES and RIXS techniques can be re-interpreted in view of the
particular distortion pattern we find.

\begin{figure}[tbp]
\includegraphics* [width=200pt] {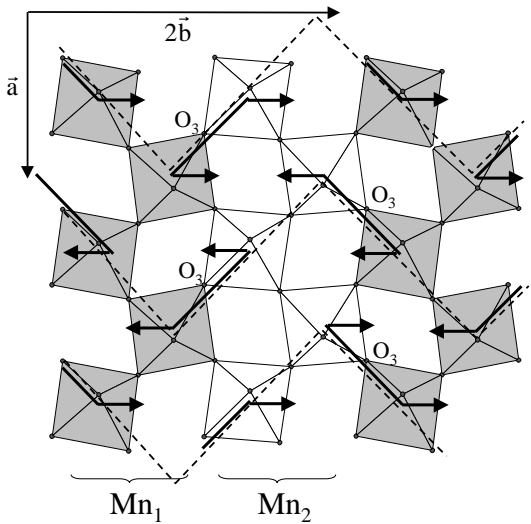}
\caption{\small  Dashed line represent the zig-zag chains of the ferromagnetically coupled spins. The
O$_3$ equatorial oxygen are shown.} \label{afce}
\end{figure}

The structure geometry at LT display elongated octahedra in the \mnomn \ direction. To understand the
covalency increase together with an electronic localization process below \tco \ evidenced by
resistivity measurements, one should imagine this increase arises within the \mnomn \ units, where one
electron is shared (delocalized) leading to mixed localized states of $d_{z^2}-2p_{\sigma}-d_{z^2}$
character. This is consistent with the \mnomn \ angle becoming the most opened angle ($161\grad$),
suggesting further this electron delocalization is favored by tuning on a local double exchange (DE)
process within each Mn$_1$-Mn$_2$ pair, maintaining the intermediate valence of both Mn atoms in the
ionic picture. In other words, this interpretation of the structural distortions emphasizes the
appearance, just below \tco, of ordered \mnomn \ molecular objects with ferromagnetically coupled Mn
moments. The hypothesis of the existence of such ferromagnetic coupled Mn-pairs gives also a clue to
interpret the anomaly in the magnetic susceptibility observed at the supposedly purely structural
transition \cite{Millange00, Lees95}, for samples that remain in a paramagnetic state below \tco.
Their formation should lead to new elemental paramagnetic units with a higher effective moment. This
seems to be confirmed by the increase of the Curie constant $C=\frac{N\mu_{eff}^2}{k_B}$ in the region
between \tco \ and \tn \ in \bisrh, where there is no magnetic contribution arising from the
paramagnetic rare-earth element \cite{Garcia-munoz01}. The non validity of the ionic picture changes
also our understanding of the CE-type magnetic structure. We think that  the GKA rules for the super
exchange interaction cannot be totally applied to the present case because the system preserves an
itinerant nature even if it is at the level of structural dimers. One should note that the
Mn$_1$-Mn$_2$ ferromagnetically coupled pairs are linked into zigzag chains, which are the building
blocks of the CE structure, as displayed in Fig.\ref{afce}. Below \tn, AF superexchange coupling can
occur between these molecular objects with localized electronic states. They should be mediated by the
non-labeled oxygen atoms of the Fig. \ref{afce} to stabilize the observed CE structure.

In conclusion, the present study invalidates the widely used ionic picture of charge ordering invoked
to interpret the jump of the resistivity and the appearance of superstructure reflections at the
structural transition in the \prca \ system. The results presented in this Letter favor a description
in term of an anisotropic charge redistribution different form the CO/OO model proposed by Goodenough.
The transition may be interpreted in terms of the condensation and ordering of ferromagnetic Mn-Mn
pairs remaining paramagnetic below \tco \ and giving rise, below \tn, to different variants of the CE
magnetic structure depending on the exact stoichiometry.

We would like to thank P.Berthet,  for his help in carrying out
the XANES experiments at LURE (ORSAY, France).

\end{document}